 \documentstyle[12pt]{article}

\let\la=\label  
  
 \def\bd{\begin{document}} \def\ed{\end{document}}
\def\ds{\documentstyle} \let\fr=\frac \let\bl=\bigl \let\br=\bigr
\let\Br=\Bigr \let\Bl=\Bigl
\let\bm=\bibitem
\let\na=\nabla
\let\pa=\partial \let\ov=\overline
\newcommand{\be}{\begin{equation}}
\newcommand{\ee}{\end{equation}}
\def\ba{\begin{array}}
\def\ea{\end{array}}
\newcommand{\ho}[1]{$\, ^{#1}$}
\newcommand{\hoch}[1]{$\, ^{#1}$}
\newcommand{\bea}{\begin{eqnarray}}
\newcommand{\eea}{\end{eqnarray}}
\newcommand{\ra}{\rightarrow}
\newcommand{\lra}{\longrightarrow}
\newcommand{\Lra}{\Leftrightarrow}
\newcommand{\ap}{\alpha^\prime}
\newcommand{\bp}{\tilde \beta^\prime}
\newcommand{\tr}{{\rm tr} }
\newcommand{\Tr}{{\rm Tr} }
\newcommand{\NP}{Nucl. Phys. }
\newcommand{\tamphys}{\it Isaac Newton Institute for Mathematical Sciences\\
University of Cambridge
 \\ 20 Clarkson Road, Cambridge CB3 OEH, U.K.}

\newcommand{\auth}{M. J. Duff \footnote{On leave of absence from the Center for
Theoretical Physics, Texas A\&M University,
College Station, Texas 77843. Research supported in part by NSF Grant
PHY-9411543.}}

\begin{document}
\hfill{NI-94-016}

\hfill{CTP-TAMU-48/94}

\hfill{hep-th/9410210}

\vspace{24pt}

\begin{center}
{ \large {\bf CLASSICAL/QUANTUM DUALITY\footnote{Talk delivered at the
International Conference
on High Energy Physics, Glasgow, July 1994.} }}

\vspace{36pt}

\auth

\vspace{10pt}

{\tamphys}

\vspace{44pt}

\underline{ABSTRACT}

\end{center}

String theory requires two kinds of loop expansion: classical $(\alpha')$
worldsheet loops with
expansion parameter $<T>$ where $T$ is a modulus field, and quantum $(\hbar)$
spacetime loops with
expansion parameter $<S>$ where $S$ is the dilaton field.  Four-dimensional
string/string duality (a
corollary of ten-dimensional string/fivebrane duality) interchanges the roles
of $S$ and $T$ and
hence interchanges classical and quantum.

{\vfill\leftline{}\vfill
\leftline{October 1994}
\pagebreak

\section{Classical/Quantum Duality}

There is now a consensus that the really important questions of string theory
will never be answered
within the framework of a weak coupling perturbation expansion. Here I describe
some recent
work which begins to address this strong coupling problem. It is based on the
idea that the same
physics may equally well be described by the fundamental four-dimensional
superstring
or by a {\it dual} four-dimensional superstring \cite{Duff1} that corresponds
to
a soliton solution\footnote{This {\it dual} string of \cite{Duff1} is not to be
confused
with the {\it stringy cosmic string} of \cite{Greene}.  The two solutions are
different.} of the
fundamental string. In this respect, the idea provides a stringy generalization
of the old
Montonen-Olive conjecture \cite{Montonen} of a duality between the electrically
charged particles of
a fundamental supersymmetric theory and its magnetically charged solitons.
Indeed,
the latter duality is in fact subsumed by the former in that the solitonic
magnetic {\it H-monopoles}
\cite{Khuri,Gauntlett} of the fundamental string are the fundamental electric
winding states of the
dual string \cite{Duff2,Duff3}.

This four-dimensional string/string duality is a corollary of the ten-
dimensional
string/fivebrane duality which states that the same physics may equally well be
described by
the fundamental ten-dimensional superstring (an extended object with one
spatial dimension) or by a
dual ten-dimensional superfivebrane \cite{Duff4} (an extended object with five
spatial
dimensions) that corresponds to a soliton solution of the fundamental string
\cite{Strominger,Lu1}.
The pay-off, if such a conjecture proves to be true, is that the strongly
coupled string corresponds
to the weakly coupled fivebrane. After compactification to four dimensions, the
fivebrane will appear
as an H-monopole or a dual string according as it wraps around $5$ or $4$ of
the compactified
dimensions  \cite{Strominger} which, for concreteness and simplicity, we take
to be a six-dimensional
torus\footnote{It could in principle also appear as a membrane by wrapping
around $3$ of the
compactified dimensions, but the fundamental four-dimensional string obtained
in this way does not
admit the membrane soliton \cite{Duff1}.}. The inverse tension of the dual
string, $2\pi \tilde
\alpha'$, is related to that of the fundamental string, $2\pi \alpha'$, by the
Dirac quantization rule
\cite{Duff1} \be 8GR^2=n\alpha'\tilde \alpha'\,\,\,\,\,\,n=integer
\la{Dirac}
\ee
where $G$ is Newton's constant and $R$ is the compactification scale. One's
first guess might
therefore be to assume that the strongly coupled four-dimensional fundamental
string corresponds to
the weakly coupled dual string, but in fact something more subtle and
interesting happens.  The
fundamental string exhibits a minimum/maximum length duality, $R\rightarrow
\alpha'/R$, called
$T$-duality, manifest order by order in perturbation theory.  There is also
evidence that it exhibits
a minimum/maximum coupling constant duality \cite{Font,Vafa}, $g\rightarrow
1/g$, called $S$-duality,
which is intrinsically non-perturbative. In going from the string to the dual
string, these two
dualities trade places leading to a {\it duality of dualities}
\cite{Schwarz,Binetruy,Duff1,Sen2} as
illustrated in Table 1.

\begin{table}
$
\begin{array}{lcc}
&Fundamental \, string&Dual \, string\\
&&\\
Moduli&T=b+ie^{-\sigma}&S=a+ie^{-\eta}\\
Worldsheet \, coupling&<e^{\sigma}>=\alpha'/R^2&<e^{\eta}>=g^2\\
Large/small \, radius &R\rightarrow \alpha'/R&g\rightarrow 1/g\\
T-duality&O(6,22;Z)&SL(2,Z)\\
Axion/dilaton&S=a+ie^{-\eta}&T=b+ie^{-\sigma}\\
Spacetime \, coupling&<e^{\eta}>=g^2&<e^{\sigma}>=\alpha'/R^2\\
Strong/weak \, coupling&g\rightarrow 1/g&R\rightarrow \alpha'/R\\
S-duality&SL(2,Z)&O(6,22;Z)
\end{array}
$
\label{Table1}
\caption{Duality of dualities}
\end{table}

String theory requires two kinds of loop expansion: classical
$(\alpha')$ worldsheet loops with expansion parameter $<e^{\sigma}>$ where
$\sigma$ is a modulus
field, and quantum $(\hbar)$ spacetime loops with expansion parameter
$<e^{\eta}>$ where $\eta$ is the
dilaton field.  Introducing the axion field $a$ and another pseudoscalar
modulus field $b$,
four-dimensional string/string duality interchanges the roles of
$S=a+ie^{-\eta}$ and
$T=b+ie^{-\sigma}$, and hence interchanges classical and quantum. Thus this
duality of dualities
exhibited by four-dimensional strings is entirely consistent with the earlier
result that
ten-dimensional string/fivebrane duality interchanges the spacetime and
worldsheet loop expansions
\cite{Lu2}, and is entirely consistent with the Dirac quantization rule
(\ref{Dirac}) that follows
from a earlier string/fivebrane rule \cite{Lu1}. Thus, for $n=1$, we have
\[
<e^{\eta}>=g^2=8G/\alpha'=\tilde \alpha'/R^2 \]
\be
<e^{\sigma}>=\tilde g^2=8G/\tilde\alpha'=\alpha'/R^2
\ee
where $\tilde g$ is the dual string spacetime loop expansion parameter.

Group theoretically, these dualities
are given by $O(6,22;Z)$ in the case of $T$-duality and $SL(2,Z)$ in the case
of $S$-duality. It has
been suggested \cite{Lu3,Duff3} that these two kinds of duality should be
united into a bigger group
$O(8,24;Z)$ which contains both as subgroups.  This would have the bizarre
effect of
eliminating the distinction between classical and quantum.

\section{Acknowledgments}
I am grateful to the Director and Staff of the Isaac Newton Institute, and to
the organizers of the
{\it Topological Defects} programme, for their hospitality.

\end{document}